\documentclass[preprint,prc,aps]{revtex4}
\usepackage{young}
\everymath={\displaystyle}
\usepackage{rotating}
\def\1{\mbox{l\hspace{-0.53em}1}}

\begin{document}
\title{Masses of $[{\bf 70},\ell^+]$ Baryons in the $1/N_c$ Expansion}

\author{N. Matagne\footnote{e-mail address: nmatagne@ulg.ac.be}}

\author{Fl. Stancu\footnote{ e-mail address: fstancu@ulg.ac.be
}}
\affiliation{University of Li\`ege, Institute of Physics B5, Sart Tilman,
B-4000 Li\`ege 1, Belgium}

\date{\today}

\begin{abstract}
The masses of positive $[{\bf 70},0^+]$ and $[{\bf 70},2^+]$ nonstrange and strange baryons
are calculated in the $1/N_c$ expansion. The approach is
based on the separation of a system of $N_c$ quarks into
an excited core and an excited quark. The previous work for two flavor baryons
is now extended to include strange baryons, 
 to  first order in SU(3)-flavor breaking.  We show that the extension
to $N_f = 3$ maintains the regularities previously observed in the
behaviour of the linear term in $N_c$, of the spin-spin and of the spin-orbit
terms. In particular the contribution of the spin-dependent terms decrease
with the excitation energy, the dominant term remaining the spin-spin term.
\end{abstract}

\maketitle

\section{Introduction}

The large $N_c$ limit of QCD  suggested by 't Hooft \cite{HOOFT}
and the power counting rules of Witten \cite{WITTEN} lead to
a consistent perturbative $1/N_c$ expansion method to study baryon spectroscopy,
which allows to compute $1/N_c$ corrections in a systematic way.
A perspective on the current research status can be found, for example, in
Ref. \cite{TRENTO}.
% Proceedings of the Workshop {\it Large $N_c$ QCD 2004}, Trento, Italy,
% July 5-11 2004, eds. J. L. Goity, R. F. Lebed, A. Pich, C. L. Schat and
% N. N. Scoccola.}.
The method is based on
the result that baryons satisfy a contracted 
spin-flavor algebra in the
large $N_c$ limit of QCD \cite{DM}, which reduces to SU(2$N_f$) for
ground state baryons,  where $N_f$ is the number of flavors. 
For $N_c \rightarrow \infty $ the baryon masses are degenerate.
At large $N_c$, the mass splitting starts at order $1/N_c$ for
the ground state baryons (N = 0 band). They belong to the
 $\bf 56$ representation of SU(6), and have been described with remarkable success
\cite{DM,DJM94,DJM95,CGO94,Jenk1,JL95,DDJM96}.
The applicability of the approach to excited states is a subject of
current investigation.
Although the SU(6) symmetry is broken for excited states, the 
experimental facts suggest a small breaking, which then
implies that the $1/N_c$ expansion can still be applied. In this case the splitting starts at 
order $N^0_c$, as we shall see below.

The excited states belonging to the $[{\bf 70},1^-]$ multiplet (N = 1 band)
have been studied extensively in SU(4) ($N_f$ = 2)
\cite{CGKM,Goi97,PY1,PY2,CCGL,CaCa98,BCCG,SCHAT,Pirjol:2003ye,cohen1}.
The approach has been extended to  $N_f$ = 3 in Ref. \cite{SGS}
and it included first order in SU(3) symmetry breaking.  There are also
a few studies of the physically important multiplets 
belonging to  the N = 2 band. These are related to
$[{\bf 56'},0^+]$  in SU(4) \cite{CC00}, to   $[{\bf 56},2^+]$  in
SU(6) \cite{GSS} and to  $[{\bf 70},\ell^+]$  in SU(4) \cite{MS2}.
The method had also been applied to highly excited nonstrange and strange 
baryons \cite{MS1} belonging to the  $[{\bf 56},4^+]$ multiplet (N = 4 band).
So far, configuration mixing has been neglected in the N = 2 band. It would involve
new parameters under the form of mixing angles which, to be well
determined, would require, generally, much more  than the existing data. 
However the power counting for configuration mixing is quite 
well established \cite{GOITY05}.

The 35 SU(6) generators are 
\begin{equation}\label{SU6}
S^i = \frac{\sigma^i }{2}  \otimes\, \1;~ T^a =\, \1 \otimes \frac{ \tau^a }{2};
~ G^{ia} = \frac{\sigma^i }{2} \otimes \frac{ \tau^a }{2},
\end{equation}
where $i=1,2,3$ and $a=1,2, \ldots , 8$. For excited states the mass operator is a linear combination of SU(2$N_f$)
and SO(3) scalars with coefficients to be determined from a fit.
They incorporate the dynamics of quarks and it is important to understand
their behaviour.  Operators which break SU(2$N_f$),
but are rotational invariant, can also be added to the
mass operator. They embed the SU(3)-flavor breaking, due to the difference
in the mass of the strange and nonstrange quarks.
The general form of an SU(6) $\times$ SO(3) scalar is
\begin{equation}\label{OLFS}
O^{n} = \frac{1}{N^{n-1}_c} O^{(k)}_{\ell} \cdot O^{(k)}_{SF},
\end{equation}
where  $O^{(k)}_{\ell}$ is a $k$-rank tensor in SO(3) and  $O^{(k)}_{SF}$
a $k$-rank tensor in SU(2), but scalar in SU(3)-flavor.
This implies that $O^{n}$ is a combination 
of SO(3) generators $\ell_i$ and of SU(6) generators (see below). 
In calculating the mass spectrum,
the general procedure is to split the baryon into an excited quark
and a core. The latter is in its ground state for the N = 1 band but
generally carries some excitation for N $>$ 1 (for example the
$[{\bf 70},\ell^+]$ multiplet \cite{MS2}). The excitation is implemented into the
orbital part of the wave function. 
The advantage of this method is that the problem is reduced to 
the known case of the ground state, because the spin-flavor part
of the core wave function remains always symmetric. But a disadvantage
is that one introduces a large number of operators of type (\ref{OLFS}).
Let us denote the excited quark operators  by $\ell^i_q$,
$s^i, t^a$ and $g^{ia}$ and the corresponding core operators by
$\ell^i_c$, $S^i_c, T^a_c$ and $G^{ia}_c$. Then, for example,
for the $[{\bf 70},1^-]$ multiplet with $N_f = 2$ one has 12 linearly independent
operators up to $1/N_c$ power included \cite{CCGL}. 

In this practice
the matrix elements of the excited quark are straightforward, as being
single-particle operators. The matrix elements of the core operators 
$S^i_c$ and $T^a_c$ are also simple to calculate, while those of $G^{ia}_c$
are more involved. 
Analytic group theory formulas for the matrix elements of all SU(4)
generators  have been derived  in Ref. \cite{HP}. They are factorized 
according to a generalized Wigner-Eckart theorem into a reduced
matrix element and an SU(4) Clebsch-Gordan coefficient. They have 
been used in nuclear physics, which is governed by the SU(4) symmetry,
but can 
be straightforwardly be applied to a system of arbitrary $N_c$ quarks 
containing the isodoublet $u, d$. Recently we have extended the approach 
of Ref. \cite{HP}  to SU(6)  \cite{MS3} and obtained matrix elements of
all SU(6) generators between symmetric $[N_c]$ states.
These matrix elements are used below.
The matrix elements of $G^{ia}_c$ with nonzero
strangeness presented in Ref. \cite{PS} are particular cases of the results
of Ref. \cite{MS3}.
%Recently, explicit formulas for SU(3) Clebsch-Gordan coefficients, relevant
%for couplings of mesons to baryons at large $N_c$, have also been derived
%\cite{CL}.

We should keep in mind that the excited states are resonances and have
a finite width. Generic large $N_c$ counting rules give widths of order
$N^0_c$ \cite{CGKM,PY1,PY2,CaCa98,cohen1,cohen2}. According to Ref.
\cite{CGKM} the narrowness of the excited states is an artifact of simple
quark model assumptions. Here, as in constituent quark models, we do
ignore the finite width and treat the resonances as bound states.

The paper is organized as follows. In the next section we recall the orbital
structure of the wave functions of the $[{\bf 70},\ell^+]$ baryon multiplet.
Section 3 is devoted to the formalism of the mass operator.
In Sec. 4 we present results for the masses of 47 nonstrange and strange 
baryons, most of which are predictions. The last section contains our
conclusions. Appendix A is devoted to the operators $O_3$, $O_4$ and to the isospin operator $O_6$
which are of order $\mathcal{O}(N_c^0)$.
The first two are operators for which the matrix elements change the analytic 
form as a function of $N_c$,  when going
from $N_f = 2$ to $N_f = 3$. Appendix B gives the general formula  for
the matrix elements of SU(3)-flavor breaking operators needed to construct $B_1$, $B_2$ and $B_4$.
Appendix C gives the matrix elements of the spin-orbit operator $O_2$.

%%%%%%%%%%%%%%%%%%%%%%%%%%%%%%%%%%%%%%%%%%%%%%%%%%%%%%%%%%%%%%%%%%%%%%%%%%%%%%% 
\section{The wave functions of $[{\bf 70},\ell^+]$ excited states}

For the time being, we adopt the usual practice and divide the system 
of $N_c$ quarks into an excited quark and a core, which can be excited or not.
%The orbital part of baryons belonging to the $[\bf 70,\ell^+]$ multiplet
%is more complicated than that of the symmetric representation $[{\bf 56},\ell^+]$
%where it is not necessary 
%to distinguish between excited and core quarks
%\cite{GSS,MS1}. It therefore requires a special treatment.  
Below we use 
the notations given in our previous work \cite{MS2}.
We introduce the quark model indices $\rho$ and 
$\lambda$  to distinguish
between the two independent orbital wave functions of the multiplet $[{\bf 70},\ell^+]$.
The first is associated with states which are antisymmetric under
the permutation of the first two particles while the second implies
symmetry under the same permutation.
Then, for $\ell = 0$ the orbital wave function is
\begin{equation}\label{L0}
|{\bf N_c-1,1}, 0^+\rangle_{\rho,\lambda} =  
\sqrt{\frac{1}{3}}|[N_c-1,1]_{\rho,\lambda}(0s)^{N_c-1}(1s)\rangle 
+\sqrt{\frac{2}{3}}|[N_c-1,1]_{\rho,\lambda}(0s)^{N_c-2}(0p)^2\rangle.
\end{equation}
In the first term $1s$ is the first (single particle)
radially excited state with $n=1$, $\ell = 0$ 
($N=2n+\ell$). In the second term     
the two quarks are excited to the $p$-shell to get $N=2$. They are coupled to 
$\ell = 0$. By analogy, for $\ell = 2$ one has
\begin{equation}\label{L2}
|{\bf N_c-1,1}, 2^+\rangle_{\rho,\lambda} =  
\sqrt{\frac{1}{3}}|[N_c-1,1]_{\rho,\lambda}(0s)^{N_c-1}(0d)\rangle
 +\sqrt{\frac{2}{3}}|[N_c-1,1]_{\rho,\lambda}(0s)^{N_c-2}(0p)^2\rangle,
\end{equation}
where the two quarks in the $p$-shell are coupled to  $\ell = 2$.
One can see that  
the coefficients of the linear combinations (\ref{L0}) and (\ref{L2})
 are independent of $N_c$
so that both terms have to be considered in the 
large $N_c$ limit. In  Eqs. (\ref{L0}) and (\ref{L2}) the first term can be 
treated as in the $[{\bf 70},1^-]$ sector, \emph{i.e.} as an excited quark
coupled to a ground state core 
\cite{CGKM,Goi97,PY1,PY2,CCGL,CaCa98,BCCG,SGS,SCHAT,Pirjol:2003ye,cohen1}.
The second term will be treated here as an excited quark coupled to an 
excited core. To see this, we rewrite it by using the fractional parentage
technique to get 
\begin{eqnarray}\label{CFP}
|[N_c-1,1]_{\rho,\lambda} (0s)^{N_c-2}(0p)^2,\ell^+ \rangle  &=& 
\sqrt{\frac{N_c-2}{N_c}} \Psi_{[N_c-1]}((0s)^{N_c-2}(0p)) \phi_{[1]}(0p) \nonumber \\
& & -\sqrt{\frac{2}{N_c}} \Psi_{[N_c-1]}((0s)^{N_c-3}(0p)^2)\phi_{[1]}(0s),
\end{eqnarray}
both for $\ell = 0$ and 2. Here all states are normalized. 
The first factor in each term in the right-hand side is a symmetric 
$(N_c-1)$-particle wave function
and $\phi_{[1]}$ is a one particle wave function associated to
the $N_c$-th particle. One can see that 
for large $N_c$ the  coefficient of the first term is $\mathcal{O}(1)$ and of the 
second $\mathcal{O}(N_c^{-1/2})$.
Then, in the large  $N_c$ limit, one can neglect the second term and take into account
only  the first term, 
where both the core and $N_c$-th particle have an $\ell = 1$ excitation.

Each of the above configurations $(0s)^{N_c-1}(1s)$, $(0s)^{N_c-1}(0d)$
or $(0s)^{N_c-2}(0p)^2$ represent  orbital parts of  a given total wave function.
We denote by $\ell_q$ and $\ell_c$ the angular momenta of the excited quark 
and of the excited core respectively. They are coupled to a total angular
momentum $\ell$. Then in SU(6) $\times$ SO(3) the most general form of the wave 
function is
\begin{eqnarray}\label{EXCORE}
\lefteqn{|\ell S;JJ_3;(\lambda \mu) Y I I_3\rangle  =%}\nonumber \\
\sum_{m_c,m_q,m_\ell,S_3}
\left(\begin{array}{cc|c}
	\ell_c    &  \ell_q   & \ell   \\
	m_c  &    m_q    & m_\ell 
      \end{array}\right) %\nonumber \\ 
   \left(\begin{array}{cc|c}
	\ell    &    S   & J   \\
	m_\ell  &    S_3  & J_3 
      \end{array}\right)}
\nonumber \\
& \times &
\sum_{p p'}   c^{[N_c-1,1]}_{p p'}(S)
|S S_3; p \rangle
|(\lambda \mu)Y I I_{3}; p'  \rangle
|\ell_qm_q\rangle   |\ell_cm_c\rangle,
\end{eqnarray}
where 
\begin{equation}
|S S_3; p \rangle = \sum_{m_1,m_2}
 \left(\begin{array}{cc|c}
	S_c    &    \frac{1}{2}   & S   \\
	m_1  &         m_2        & S_3
      \end{array}\right)
      |S_cm_1 \rangle |1/2m_2 \rangle,
\end{equation}
with $p = 1$ if  $S_c = S - 1/2$ and $p = 2$ if $S_c = S + 1/2$ and
\begin{eqnarray}\label{statessu(3)}
|(\lambda \mu)Y I I_{3}; p' \rangle  =
\sum_{Y_c,I_c,I_{c_3},y,i,i_{3}}
\left(\begin{array}{cc|c}
	(\lambda_c \mu_c)    &   (10)  & (\lambda \mu) \\
	 Y_cI_cI_{c_3} &   y i i_{3} &  YII_{3}
      \end{array}\right) %\nonumber \\
|(\lambda_c \mu_c) Y_cI_cI_{c_3}\rangle
|(10) y i i_{3} \rangle,
\end{eqnarray}
where $p' = 1$ if $(\lambda_c \mu_c) = (\lambda - 1, \mu)$,
$p' = 2$ if $(\lambda_c \mu_c) = (\lambda + 1, \mu - 1)$ and $p' = 3$ if $(\lambda_c \mu_c) = (\lambda, \mu + 1)$.
The spin-flavor  part of the wave function (\ref{EXCORE})
of symmetry $[N_c-1,1]$ results from the inner product of the 
spin and flavor wave functions.
The indices $p$ and $p'$ represent the row where the last particle
(the excited quark) is located in the Young diagram of SU(2)-spin
and SU(3)-flavor states respectively. Thus the  
coefficients  $c^{[N_c-1,1]}_{p p'}(S)$ are  isoscalar factors 
\cite{book,ISOSC} of the 
permutation group of ${N_c}$ particles, the expressions of which are \cite{MS3}
\begin{eqnarray}\label{SU2}
c^{[N_c-1,1]}_{11}(S) & =  & - \sqrt{\frac{(S + 1)(N_c - 2 S)}{N_c(2 S + 1)}},
\nonumber \\
c^{[N_c-1,1]}_{22}(S) &  = & \sqrt{\frac{S[N_c+2(S + 1)]}{N_c(2 S + 1)}},
\nonumber \\
c^{[N_c-1,1]}_{12}(S) &  = & c^{[N_c-1,1]}_{21}(S) = 1,
\nonumber \\
c^{[N_c-1,1]}_{13}(S) &  = & 1.
\end{eqnarray} 
In Eqs. (\ref{decuplet2})-(\ref{singlet2})  below, we illustrate their
application for $N_c$ = 7. In each inner product 
the first Young diagram corresponds to spin and the second to flavor. Accordingly,
one can see that Eq. (\ref{decuplet2}) stands for $^210$, Eq. (\ref{octet4}) 
for $^48$, Eq. (\ref{octet2}) for $^28$ and Eq. (\ref{singlet2}) for $^21$.
Each inner product contains the corresponding isoscalar factors and 
the position of the last particle is marked with a cross. In the right hand side, from the location of the cross one can read off the values of $p$ and of $p'$.
The equations are 
\begin{eqnarray}
\label{decuplet2}
\raisebox{-9.0pt}{\mbox{\begin{Young}
 & & & & & \cr
$\times$ \cr
%\cr
\end{Young}}}\
& = &
c^{[6,1]}_{21}\! \! \!
\raisebox{-9.0pt}{\mbox{
\begin{Young}
& & & \cr
& & $\times$\cr
\end{Young}}} \ \times \! \! \! \! \!
\raisebox{-9.0pt}{\mbox{
\begin{Young}
& & & & $\times$\cr
& \cr
\end{Young}}}\ ,
\\ \nonumber
\\
\label{octet4}
\raisebox{-9.0pt}{\mbox{\begin{Young}
 & & & & & \cr
$\times$ \cr
%\cr
\end{Young}}}\
& = &
c^{[6,1]}_{12}\! \! \!
\raisebox{-9.0pt}{\mbox{
\begin{Young}
& & & & $\times$\cr
& \cr
\end{Young}}} \ \times \! \! \! \! \!
\raisebox{-9.0pt}{\mbox{
\begin{Young}
& & & \cr
& & $\times$\cr
\end{Young}}}\ ,
\\ \nonumber
\\
\label{octet2}
\raisebox{-9.0pt}{\mbox{\begin{Young}
 & & & & & \cr
$\times$ \cr
%\cr
\end{Young}}}\
&=& c^{[6,1]}_{11}\! \! \!
\raisebox{-9.0pt}{\mbox{
\begin{Young}
& & & $\times$\cr
& & \cr
\end{Young}}} \ \times \! \! \! \! \!
\raisebox{-9.0pt}{\mbox{
\begin{Young}
& & & $\times$\cr
& & \cr
\end{Young}}} \nonumber \\
& & + \ c^{[6,1]}_{22}\! \! \!
\raisebox{-9.0pt}{\mbox{
\begin{Young}
& & & \cr
& & $\times$\cr
\end{Young}}} \ \times \! \! \! \! \!
\raisebox{-9.0pt}{\mbox{
\begin{Young}
& & & \cr
& & $\times$\cr
\end{Young}}}\ ,
\\ \nonumber
\\
\label{singlet2}
\raisebox{-9.0pt}{\mbox{\begin{Young}
 & & & & & \cr
$\times$ \cr
%\cr
\end{Young}}}\
&=& c^{[6,1]}_{13}\! \! \!
\raisebox{-9.0pt}{\mbox{
\begin{Young}
& & & $\times$\cr
& & \cr
\end{Young}}} \ \times \! \! \! \! \!
\raisebox{-15pt}{\mbox{
\begin{Young}
& & \cr
& & \cr
$\times$ \cr
\end{Young}}}\ .
\end{eqnarray}
For the configurations $(0s)^{N_c-1}(1s)$ and  $(0s)^{N_c-1}(0d)$
the expression (\ref{EXCORE}) slightly simplifies because $\ell_c = 0$. Only 
for the configuration $(0s)^{N_c-2}(0p)^2$ the core is excited with $\ell_c = 1$,
in agreement with the discussion following Eq. (\ref{CFP}).
%%%%%%%%%%%%%%%%%%%%%%%%%%%%%%%%%%%%%%%%%%%%%%%%%%%%%%%%%%

\section{The mass operator}

For the $[{\bf 70},\ell^+]$ multiplet the mass operator
can be written as the linear combination 
\begin{equation}
\label{massoperator}
M_{[{\bf 70},\ell^+]} = \sum_{i=1}^6 c_i O_i + d_1 B_1 + d_2 B_2 + d_4B_4,
\end{equation} 
where the operators $O_i$ are of type (\ref{OLFS}) and $B_i$ are 
SU(6) breaking operators defined below.
The values of the coefficients $c_i$ and $d_i$
which encode the QCD dynamics, are given in Table \ref{operators}.
They were found by a numerical fit described in the next section.

The building blocks of  $O_i$ and  $B_i$
are the excited core operators $\ell^i_c$, 
$S^i_c$, $T^a_c$ and $G^{ia}_c$ and the excited quark operators $\ell^i_q$, 
$s^i$, $t^a$ and $g^{ia}$. We also introduce the rank $k=2$ tensor
operator \footnote{The irreducible spherical tensors are defined according to
Ref. \cite{BRINK}.}
\begin{equation}\label{TENSOR}
\ell^{(2),ij}_{ab}=\frac{1}{2}\left\{\ell^i_a,\ell^j_b\right\}-\frac{1}{3}\delta_{i,-j}\vec{\ell}_a\cdot\vec{\ell}_b,
\end{equation}
with $a=c$, $b=q$ or vice versa or $a=b=c$ or $a=b=q$. For simplicity 
when $a=b$, we use a single index $c$, for the core, or $q$ for the 
excited quark so that the tensor operators become $\ell^{(2),ij}_c$ and
$\ell^{(2),ij}_q$ respectively. The latter case represents the tensor
operator used in the analysis of the $[{\bf 70},1^-]$ multiplet (see \emph{e.g.} 
Ref. \cite{CCGL}).

There are many linearly independent operators $O_i$ and $B_i$ 
which can be constructed from the excited quark and the core operators.  
Here, due to lack of data,
we have considered a restricted list containing the most dominant
operators in the mass formula. The selection was determined from the
previous experience of Refs. \cite{CCGL} and \cite{MS2} for $N_f = 2$
and of Ref. \cite{SGS} for  $N_f$ = 3. The 
operators $O_i$ entering Eq. (\ref{massoperator}) are listed
in Table \ref{operators}. $O_1$ is linear in $N_c$ and is the
most dominant in the mass formula. At $N_c \rightarrow \infty $
is the only one which survives. $O_2$ is the dominant part 
of the spin-orbit operator. It acts on the excited quark and is of 
order $N^0_c$. The operator $O_3$ is a composite two-body operator.
It contains the tensor operator (\ref{TENSOR}) which acts on the 
excited quark and the generators $g^{ia}$ and $G^{ja}_c$ acting on the
the excited quark and on the core respectively. 
The contribution of $G^{ja}_c$ sums coherently, thus it
introduces an extra power in $N_c$,
which implies that the matrix elements $O_3$ are of order $N^0_c$. 
For the same reason the matrix elements of  
$O_4$ are also of order  $N^0_c$. 
As explained in the next section, we could not obtain its coefficient $c_4$,
because of  scarcity of data for the $[{\bf 70},\ell^+]$ multiplet.  
The spin-spin operator $O_5$ is of order $1/N_c$, but its
contribution dominates over all the other terms of the mass operator
containing spin.

Here we take into account the isospin-isospin operator, denoted by $O_6$,
having matrix elements of order $N^0_c$ due to the
presence of $T_c$ which sums coherently. Up to a subtracting constant,
it is one of the four independent operators of order $N^0_c$,
which, together with $O_1$, are needed to describe the 
submultiplet structure of $[{\bf 70},1^-]$ \cite{COLEB}. Incidently, this operator 
has been omitted in the analysis of Ref. \cite{SGS}. Its coefficent $c_6$ is 
indicated in Table \ref{operators}.
 
In Tables \ref{NUCLEON}, \ref{DELTA} and \ref{SINGLET}
we show the diagonal matrix elements
of  the operators $O_i$ for octet, decuplet and flavor singlet states 
respectively. From these tables one can obtain the large $N_c$ mentioned 
above. Details about $O_3$ are given in Appendix A. 
Its matrix elements  change the analytic dependence on $N_c$ 
in going from SU(2) to SU(3). This happens for octet resonances 
which can be seen 
by comparing the column 3 of Table \ref{NUCLEON} with the corresponding
result from Ref. \cite{MS2}. The change is that the factor $N_c + 1$ in SU(2)
becomes $N_c + 1/3$ in SU(3). The same change takes place for all operators
$O_i$ containing $G^{ja}_c$ as for example the operator $O_4$ also
presented in Appendix A.

The SU(6) breaking operators, $B_1$ and $B_2$ and $B_4$ in the notation 
of Ref. \cite{SGS},
expected to contribute to the mass are listed in Table \ref{operators}.
The operators $B_1$, $B_2$ are the standard breaking operators 
while $B_4$ is directly related to the spin-orbit splitting.
They break the SU(3)-flavor symmetry to first order in $\epsilon \simeq 0.3$ 
where $\epsilon$ is proportional to the mass difference between the strange 
and $u, d$ quarks.
Table  %\ref{T8} 
V gives the matrix elements of the excited quark operator 
$t_8$ and of the core operator $T^c_8$ which are necessary to construct
the matrix elements of $B_1$ and $B_2$.
These expressions have been obtained as indicated in Appendix B.
It is interesting to note that they are somewhat different from those  
of Ref. \cite{SGS}. However for all cases with physical quantum numbers but any $N_c$, our values are
identical to those of  Ref. \cite{SGS}, so that for $N_c = 3$ there is
no difference.

For completeness, Table \ref{b3} gives the matrix elements 
of $3\ell^ig^{i8}$ needed to construct $B_4$. 
They were obtained from the formula (\ref{B4}) derived in Appendix B. As above, they are different from these of Ref. \cite{SGS} except for physical quantum numbers.
Unfortunately none of  
the presently known resonances has nonvanishing matrix elements for $B_4$.
By definition all $B_i$ have zero matrix elements for nonstrange resonances.
In addition, the matrix elements of $B_4$ for $\ell$ = 0 resonances also cancel
and for the two remaining experimentally known strange resonances they also
cancel out.  
For this reason the coefficient $d_4$ could not be determined.
%%%%%%%%%%%%%%%%%%%%%%%%%%%%%%%%%%%%%%%%%%%%%%%%%%%%%%%%%%

\section{Results}

Comparing Table I with our previous
results Ref. \cite{MS2} for nonstrange baryons, one can see that the
addition of strange baryons in the fit 
have not much changed  the values of the coefficients $c_1$ and $c_5$
(previously $c_4$).
The spin-orbit coefficient $c_2$
had changed sign but in absolute value remains small.
The resonance $F_{05} (2100)$ is mostly responsible for this change.
But actually the crucial experimental input for the spin-orbit contribution
should come from $\Lambda$'s, as in the case of the 
$[{\bf 70},1^-]$ multiplet \cite{SGS}. Unfortunately data for 
the two flavor singlets with $\ell \neq 0$,
$^2\Lambda'[{\bf 70},2^+]5/2$ and $^2\Lambda'[{\bf 70},2^+]3/2$, 
which are spin-orbit partners are missing (see Table \ref{MASSES}).
If observed, they
 will help to fix the strength and sign of 
the spin-orbit terms unambiguously inasmuch as $O_3$, $O_4$ and $O_5$ do 
not contribute to their mass.

Presently,  due to the large uncertainty obtained from the fit of $c_2$,
there is still some overlap with the  value obtained from nonstrange 
resonances. 
The coefficient $c_3$ is about twice smaller in absolute value now. Interestingly,
the present values of the coefficients $c_1$, $c_2$  and $c_5$ follow the
trend discussed in Ref. \cite{MS2}, namely the spin-spin
and the spin-orbit contributions decrease with the excitation energy,
 the dominant part remaining the spin-spin term,
similar to constituent quark model results
with a hyperfine interaction. 

Regarding the SU(3) breaking terms, the coefficient
$d_1$ is has opposite sign as compared to that of Ref. \cite{SGS} and
is about four times larger in absolute value.
 The coefficient $d_2$ has the same sign and about the same order 
 of magnitude. One can conclude that the 
SU(3)-flavor breaking is roughly similar in the $[{\bf 70},1^-]$ and the 
$[{\bf 70},\ell^+]$ multiplets.

The resonances belonging to the $[\bf 70,\ell^+]$ together with their
calculated masses are presented in Table \ref{MASSES}. 
The angular momentum coupling allows for 8 octets, with $J$ ranging from
7/2 to 1/2, three decuplets with $J$ from 5/2 to 1/2 and three flavor singlets
with $J$ = 5/2, 3/2 or 1/2. Ignoring isospin breaking, there are in all 47 
resonances
from which 12 are fitted and 35 are predictions. 
The best fit gave $\chi^2_{\rm dof} \simeq 1$.
Among the presently 12 resonances only five are new, the strange resonances.
This reflects the fact that the experimental situation is still
rather poor in this energy range. The known resonances are three-, two- and one-star. 

For all masses the main contribution comes from the operator $O_1$.
In the context of a constituent quark model this corresponds to the
contribution of the spin-independent part of the Hamiltonian, namely
the free mass term plus the kinetic and the confinement energy. A difference 
is that,  this contribution is constant for all resonances here, 
while in quark models the mass difference between the strange and the $u, d$ 
quarks is taken into account explicitly in the free mass term. Here 
this difference is embedded into the flavor breaking terms $B_i$  .
 
The spin-orbit operator $O_2$ naturally contributes to states with $\ell \neq 0$ only.
The operator $O_3$ contributes to states with $S = 3/2$ only. For $S=1/2$ states it gives no contribution either due to the cancellation of a $6j$ coefficient or when the wave function has $S_c=0$, as for example for flavor singlet states.

We have analyzed the role of the operator $O_4$ described in 
Appendix B. This is an operator of order $N^0_c$, like
$O_2$, $O_3$ and $O_6$. 
As in Refs. \cite{CCGL} and \cite{SGS}, 
the combination $O_2 + O_4$ is of order $1/N_c$ for octets and decuplets,
but this is no longer valid for flavor singlets. It means that the operators 
$O_2$ and $O_4$ are independent in SU(3) and both have to be included in the fit. 
However, the inclusion of $O_4$ considerably deteriorated the fit,
by abnormally increasing the spin-orbit contribution with one order of
magnitude. Therefore the contribution of  $O_4$ 
cannot be constrained with the present data and we have to wait 
until more data will be available, especially on strange resonances.

To estimate the role of the isospin-isospin operator $O_6$ we have made 
a fit without the contribution of this operator. This fit gave 
$\chi^2_{\rm dof} \simeq 0.9$ and about the same values for
$c_i$ and $d_i$ as that with $O_6$ included. This means that the 
presence of $O_6$ is not essential at the present stage.

The fitted value of the $N(1990) F_{17}$ resonance slightly deteriorates
with respect to the SU(4) case. The reason is the negative contribution 
of the spin-orbit term. Further analysis, based on more data, is 
needed in the future, to clarify the change of sign in the spin-orbit term.

Of special interest is the fact that the resonance $\Lambda(1810) P_{01}$
gives the best fit when interpreted as a flavor singlet. Such an interpretation
is in agreement with that of Ref. \cite{GR} where the baryon spectra were
derived from a flavor-spin hyperfine interaction, rooted in
pseudo-scalar meson (Goldstone boson) exchange. Thus the 
flavor-spin symmetry is common to both calculations. Moreover, the
dynamical origin of the operator $O_3$, which does not directly contribute
to $\Lambda(1810) P_{01}$, but plays an important role
in the total fit, is thought to be related to pseudo-scalar meson exchange
\cite{CCGL}. Hopefully, this study may help 
in shedding some light on the QCD dynamics hidden in the coefficients
$c_i$.

%%%%%%%%%%%%%%%%%%%%%%%%%%%%%%%%%%%%%%%%%%%%%%%%%%%%%%%%%%%%%%%%%%%%%%%%%%

\section{Conclusions}
The present results confirm the behaviour  of some of the coefficients $c_i$ of the
mass formula at large excitation energy, observed previously \cite{MS2}.
This shows that the importance of spin-dependent terms of the mass operators
vanish with the excitation energy. At any energy, these terms are dominated 
by the spin-spin contribution, like in constituent quark model studies. 
Thus the $1/N_c$ expansion can provide a deeper understanding of the successes 
of the quark models.

We have also found that the SU(3) breaking corrections are comparable in
size with the $1/N_c$ corrections, as for the 
$[{\bf 70},1^-]$ multiplet \cite{SGS} which successfully explained the
$\Lambda(1520) - \Lambda(1405)$ splitting.

The analysis of the $[{\bf 70}, \ell^+]$ remains an open problem. It  depends
on future experimental data which may help to clarify the role of
 various terms contributing to the mass operator and in particular of
$O_2$ and $O_4$. The present approach provides the theoretical
framework to pursue this study.

%%%%%%%%%%%%%%%%%%%%%%%%%%%%%%%%%%%%%%%%%%%%%%%%%%%%%%%%%%%%%%%%%%%%%%%%%%

\appendix
\section{The operators $O_3$ and $O_4$}
Here we derive analytic expressions for the matrix elements of the operators
$O_3$ and $O_4$ in SU(6). The compact form of $O_3$ is given in Table 
\ref{operators}
\begin{equation}\label{OPO3}
O_3 = \frac{3}{N_c}\ell^{(2),ij}_{q}g^{ia}G_c^{ja}.
\end{equation}
Writing the scalar products in an explicit form we have
\begin{equation}
O_3 = \frac{3}{N_c}\sum_{ij}(-1)^{i+j}\ell^{(2),-i,-j}_{q}
\sum_{Y^a I_3^a}(-1)^{I_3^a+Y^a/2}g^{ia}G_c^{ja},
\end{equation}
with $i,j = 1,2,3$ and $a = 1,2, ...,8$ and where $\ell^{(2)ij}_{q}$ is defined in Eq. (\ref{TENSOR}).
The matrix elements of the tensor operator are
\begin{equation}
\langle \ell' m' | \ell^{(2),ij} | \ell m \rangle = \delta_{\ell\ell'}
\left[ \frac{\ell (\ell+1)(2 \ell-1)(2 \ell+3)}{6}\right]^{1/2}
\sum_{\mu}   
\left(\begin{array}{cc|c}
	1 & 1 & 2   \\
	i & j & \mu
	 \end{array}\right)  
\left(\begin{array}{cc|c}
	\ell & 2 & \ell'   \\
	 m   & \mu & m' 
	 \end{array}\right).
\end{equation}
The other basic ingredients are the matrix elements of the operators 
$g^{ia}$ and $G_c^{ja}$. As explained in Ref \cite{MS3}, we have
\begin{equation}
\langle \frac{1}{2} m_2;(10) y' i' i'_3 |g^{ia}| \frac{1}{2} m_2; (10) y i i_3 \rangle =
 \left(\begin{array}{cc|c}
	\frac{1}{2} & 1 & \frac{1}{2}   \\
	     m_2     & i & m'_2
	 \end{array}\right)
 \left(\begin{array}{cc|c}
	(10)      &   (11)  & (10) \\
	 y i i_3 &     y^ai^ai^a_3 &  y' i' i'_{3}
      \end{array}\right),
     \end{equation}
and
\begin{eqnarray}\label{GIA}
\lefteqn{\langle [N_c-1] S_c'm_1';(\lambda'_c\mu'_c) Y'_cI'_cI'_{c_3}|G^{ja}_c|[N_c-1] S_cm_1;(\lambda_c\mu_c) Y_cI_cI_{c_3}\rangle = } \nonumber \\
& & \frac{1}{\sqrt{2}} \sqrt{\frac{5}{12}(N_c-1)(N_c+5)} \left(\begin{array}{cc|c}
	S_c & 1 & S_c'   \\
	   m_1  & j & m'_1
	 \end{array}\right)
\left(\begin{array}{cc|c}
	I_c & I^a & I_c'   \\
        I_{c_3}  & I^a_3 & I'_{c_3}
	 \end{array}\right)\nonumber \\
& &\times \sum_{\rho = 1,2}
 \left(\begin{array}{cc||c}
	(\lambda_c \mu_c)    &  (11)   &   (\lambda'_c \mu'_c)\\
	Y_c I_c   &  Y^a I^a  &  Y'_c I'_c
      \end{array}\right)_{\rho}
\left(\begin{array}{cc||c}
	[N_c-1]    &  [21^4]   & [N_c-1]   \\
	(\lambda_c \mu_c) S_c  &  (11)1  &  (\lambda'_c \mu'_c) S'_c
      \end{array}\right)_{\rho},
\end{eqnarray}
where the SU(3) isoscalar factors are from Ref. \cite{HECHT} and the SU(6) isoscalar factors can be found in Table 1 of Ref. \cite{MS3}.
The final formula for the matrix elements of $O_3$ between states of
mixed orbital symmetry  $[N_c-1,1]$ is
\begin{eqnarray}
\label{O3}
\lefteqn{\langle  \ell S'; JJ_3; (\lambda \mu) Y I I_3
| O_3 | \ell S; JJ_3;
(\lambda \mu) Y I I_3 \rangle = } \nonumber \\
& & (-1)^{\ell_q+\ell_c+S'+J+1} \frac{5}{4}(2 \ell+1)
 \sqrt{\ell_q (\ell_q+1)(2 \ell_q-1)(2 \ell_q+1)(2 \ell_q+3)}
 \left\{\begin{array}{ccc}
	\ell     &  2       &  \ell  \\
	\ell_q &  \ell_c  & \ell_q 
      \end{array}\right\} 
 \nonumber \\
 & & \times
 \sqrt{2(N_c-1)(N_c+5)(2S+1)(2S'+1)}
    \left\{\begin{array}{ccc}
	S    &   2   & S'   \\
	\ell    &   J   & \ell
      \end{array}\right\} 
    \nonumber \\
& &\times
       \sum_{p,p',q,q'} c^{[N_c-1,1]}_{pp'}(S)    c^{[N_c-1,1]}_{qq'}(S')\sqrt{2S'_c+1}
       \left\{\begin{array}{ccc}
	S'_c    &  S'  &  1/2  \\
	1       &  2   &  1  \\
	S_c     &  S   &  1/2
      \end{array}\right\}
      \nonumber \\
& &\times\sum_{\rho=1,2}
      U((\lambda_c \mu_c)(11)(\lambda \mu)(10);(\lambda'_c \mu'_c)(10))_{\rho}
       \left(\begin{array}{cc||c}
	[N_c-1]    &   [21^4]   &  [N_c-1]   \\
	(\lambda_c \mu_c) S_c   &   (11) 1   &  (\lambda'_c \mu'_c) S'_c
      \end{array}\right)_{\rho},
\end{eqnarray} 
where the coefficients $c^{[N_c-1,1]}_{pp'}(S)$ are given by Eqs. (\ref{SU2}).
We recall that $S_c=S-1/2$ for $p=1$ and $S_c=S+1/2$ for $p=2$
and by analogy $S'_c=S'-1/2$ for $q=1$ and $S'_c=S'+1/2$ for $q=2$.
Also $(\lambda_c \mu_c) = (\lambda-1,\mu)$ for $p'=1$,
$(\lambda_c \mu_c) = (\lambda+1,\mu-1)$ for $p'=2$ and $(\lambda_c \mu_c) = (\lambda,\mu+1)$ for $p'=3$ and an analogous
situation for $(\lambda'_c \mu'_c)= (\lambda-1,\mu)$ if $q'=1$,
 $(\lambda'_c \mu'_c)=(\lambda+1,\mu-1)$ if $q'=2$ and $(\lambda'_c \mu'_c)= (\lambda,\mu+1)$ if $q'=3$.

When applied on the excited quark the operator $O_4$ reads \cite{SGS}
\begin{equation}\label{OPO4}
O_4 = \frac{4}{N_c+1}\ell^i_{q}t^{a}G_c^{ia}.
\end{equation}
Writing the scalar products explicitly we have
\begin{equation}
O_4 = \frac{4}{N_c+1}\sum_{i}(-1)^{i}\ell^{i}_{q}
\sum_{Y^a I_3^a}(-1)^{I_3^a+Y^a/2}t^{-a}G_c^{-ia}~.
\end{equation}
The matrix elements of $t^{a}$ are \cite{MS3}
\begin{eqnarray}\label{TA}
\langle (10)y'i'i'_3
|t^{-a}| (10)yii_3\rangle
= \sqrt{\frac{4}{3}}
  \left(\begin{array}{cc|c}
	i  &    I^a    &   i'   \\
        i_3 &    -I^a_3  &   i'_3
      \end{array}\right)
\left(\begin{array}{cc||c}
(10)  &  (11)   &  (10)  \\
yi & -Y^aI^a & y'i' \\
\end{array}\right)
\end{eqnarray}
Inserting the above expression and the matrix elements of $G^{ia}_c$
, Eq. (\ref{GIA}), into (\ref{OPO4}) one obtains
\begin{eqnarray}
\label{O4}
\lefteqn{\langle  \ell S'; JJ_3; (\lambda \mu) Y I I_3
| O_4 | \ell S; JJ_3;
(\lambda \mu) Y I I_3 \rangle = (-1)^{\ell_q+\ell_c+S'-S+J+1/2} \frac{4}{N_c+1}
} \nonumber \\
& & \times
(2 \ell+1)
 \sqrt{\ell_q (\ell_q+1)(2 \ell_q+1)}
 \left\{\begin{array}{ccc}
	\ell     &  1       &  \ell  \\
	\ell_q &  \ell_c  & \ell_q 
      \end{array}\right\} 
 \nonumber \\
 & & \times
 \sqrt{\frac{5}{18} (N_c-1)(N_c+5)(2S+1)(2S'+1)}
    \left\{\begin{array}{ccc}
	J   &   \ell   & S   \\
	1   &   S'   & \ell
      \end{array}\right\} 
    \nonumber \\
& &\times
       \sum_{p,p',q,q'} c^{[N_c-1,1]}_{pp'}(S) c^{[N_c-1,1]}_{qq'}(S')\sqrt{2S'_c+1}(-1)^{-S_c'}
       \left\{\begin{array}{ccc}
	S'  &  1/2  & S'_c   \\
	S_c &  1    & S
      \end{array}\right\}
      \nonumber \\
& &\times\sum_{\rho=1,2}
      U((\lambda_c \mu_c)(11)(\lambda \mu)(10);(\lambda'_c \mu'_c)(10))_{\rho}
       \left(\begin{array}{cc||c}
	[N_c-1]    &   [21^4]   &  [N_c-1]   \\
	(\lambda_c \mu_c) S_c   &   (11) 1   &  (\lambda'_c \mu'_c) S'_c
      \end{array}\right)_{\rho}.
\end{eqnarray} 
The unitary Racah coefficients $U$, defined according to Ref.
\cite{HECHT}, which are needed to calculate (\ref{O3}) and (\ref{O4}) have
been obtained as in Ref. \cite{MS3}. Their explicit
forms are
\begin{eqnarray}
U((\lambda-1,\mu)(11)(\lambda\mu)(10);(\lambda+1,\mu-1)(10))& = &
-\frac{1}{2}\sqrt{\frac{3(\lambda+2)\mu}{2(\lambda+1)(\mu+1)}},
\nonumber \\
U((\lambda+1,\mu-1)(11)(\lambda\mu)(10);(\lambda-1,\mu)(10))& = &
\frac{1}{2}\sqrt{\frac{3\lambda(\lambda+\mu+1)}{2(\lambda+1)(\lambda+\mu+2)}}, \nonumber \\
U((\lambda,\mu +1)(11)(\lambda\mu)(10);(\lambda,\mu +1)(10))_{\rho=1} & = & \frac{\lambda+2\mu+8}{4\sqrt{g_{\lambda,\mu+1}}}, \nonumber \\
U((\lambda,\mu +1)(11)(\lambda\mu)(10);(\lambda,\mu +1)(10))_{\rho=2} & = & \frac{1}{4} \sqrt{\frac{3\lambda(\lambda+2)(\mu+3)(\lambda+\mu+4)}{(\mu+1)(\lambda+\mu+2)g_{\lambda,\mu+1}}}, \nonumber \\
U((\lambda-1,\mu)(11)(\lambda \mu)(10);(\lambda-1, \mu)(10))_{\rho=1}
&  = & - \frac{2 \lambda + \mu - 2}
 { 4 \sqrt{g_{\lambda-1,\mu}}},
 \nonumber \\
 U((\lambda-1,\mu)(11)(\lambda \mu)(10);(\lambda-1, \mu)(10))_{\rho=2}
& = & \frac{1}{4} % \nonumber \\
 \sqrt{\frac{3(\lambda + \mu)(\lambda - 1) \mu (\mu + 2)}
{(\lambda + 1)(\lambda + \mu + 2) g_{\lambda-1,\mu}}},
\nonumber \\
U((\lambda+1,\mu-1)(11)(\lambda \mu)(10);(\lambda+1, \mu-1)(10))_{\rho=1}
& = & \frac{\lambda - \mu + 5}{4 \sqrt{g_{\lambda+1,\mu-1}}},
 \nonumber \\
U((\lambda+1,\mu-1)(11)(\lambda \mu)(10);(\lambda+1, \mu-1)(10))_{\rho=2}
& = &   \nonumber \\  - \frac{1}{4}
\sqrt{\frac{3(\lambda + \mu + 1)(\lambda + \mu + 3)(\lambda + 3)(\mu - 1)}
{(\lambda + 1)(\mu + 1) g_{\lambda+1,\mu-1}}}, & &
  \end{eqnarray} 
  where    
  \begin{equation}\label{CSU3}
g_{\lambda\mu}= {\lambda}^2+{\mu}^2+\lambda\mu+3\lambda+3\mu.
\end{equation}
All $U$ coefficients, but the 4th one, are of order $\mathcal{O}(N_c^0)$ which can 
be seen by inserting $\lambda=2S$ and $\mu=N_c/2-S$. This helps in finding the 
order of the matrix elements of $O_4$.

The matrix elements of $O_6$ are given by the product of $1/N_c$ and
\begin{eqnarray}\label{HARD}
\langle \ell S JJ_3;(\lambda'\mu') Y' I' I'_3|t^aT^a_c|\ell S JJ_3;(\lambda \mu) Y I I_3\rangle = \delta_{\lambda\lambda'}\delta{\mu\mu'}\delta_{YY'}\delta_{II'}\delta_{I_3I_3'} \nonumber \\
\times (-1) \sum_{pp'} \left[c^{[N_c-1,1]}_{pp'}(S)\right]^2 \frac{2\sqrt{g_{\lambda_c\mu_c}}}{3} U((\lambda_c\mu_c)(11)(\lambda\mu)(10);(\lambda_c\mu_c)(10))_1
\end{eqnarray}
where the (-1) sign results from a phase entering the symmetry property of
SU(3) Clebsch-Gordan coefficients \cite{DESWART}. This is

\begin{equation}
 \left(\begin{array}{cc||c}
	(10)    &  (11)   &   (10)\\
	Y I   &  Y^a I^a  &  Y'I'
      \end{array}\right)
= \xi_1 (-1)^{I+I^a-I'}
\left(\begin{array}{cc||c}
	(11)    &  (10)   &   (10)\\
	Y^aI^a   &  Y I  &  Y'I'
      \end{array}\right).
\end{equation}
where $\xi_1=-1$ in this case.  The same property has also been used
in the calculation of the matrix elements of 
$O_3$ and $O_4$. A simpler alternative is to calculate the matrix elements 
of $O_6$ by using the identity

$$t\cdot T_c = \frac{1}{2}(T^2-T^2_c-t^2)$$

which gives
\begin{equation}\label{EASY}
\langle \ell S JJ_3;(\lambda\mu) Y I I_3|t^aT^a_c|\ell S JJ_3;(\lambda \mu) Y I I_3\rangle 
= \frac{1}{6} \left\{ g_{\lambda\mu}
- \sum_{pp'} \left[c^{[N_c-1,1]}_{pp'}(S)\right]^2 g_{\lambda_c\mu_c}- 4 \right\}
\end{equation}

The formulas (\ref{HARD}) and (\ref{EASY}) give identical results.

%%%%%%%%%%%%%%%%%%%%%%%%%%%%%%%%%%%%%%%%%%%%%%%%%%%%%%%%%%%%%%%%%%%%%%%%%
\section{}

Here we reproduce the general formulas 
\cite{MS3} of the matrix elements 
of the flavor breaking operators $t_8$, $T^c_8$ and $\ell_q^ig^{i8}$ which have been used 
to generate Table \ref{T8} and \ref{b3}. These are
\begin{eqnarray}
\langle \ell S JJ_3;(\lambda'\mu') Y' I' I'_3|T^8_c|\ell S JJ_3;(\lambda \mu) Y I I_3\rangle =%\nonumber \\
 \delta_{YY'}\delta_{II'}\delta_{I_3I_3'}\sum_{p,p',p''}c_{pp'}(S)c_{p p''}(S) \nonumber \\
\times \sum_{Y_c,I_c,y,i} \frac{3Y_c}{2\sqrt{3}}
\left(
\begin{array}{cc||c}
(\lambda_c\mu_c) & (10) & (\lambda\mu) \\
Y_cI_c  & yi  & YI \\
\end{array}
\right)
\left(
\begin{array}{cc||c}
(\lambda_c\mu_c) & (10) & (\lambda'\mu') \\
Y_cI_c  & yi  & YI
\end{array}
\right),
\end{eqnarray}

\begin{eqnarray}
\langle \ell S JJ_3;(\lambda'\mu') Y' I' I'_3|t^8|\ell S JJ_3(\lambda \mu) Y I I_3\rangle =%} \nonumber \\
\delta_{YY'}\delta_{II'}\delta_{I_3I_3'}\sum_{p,p',p''}c_{pp'}(S)c_{p p''}(S) \nonumber \\
\times \sum_{Y_c,I_c,y,i}
\frac{3y}{2\sqrt{3}}
\left(
\begin{array}{cc||c}
(\lambda_c\mu_c) & (10) & (\lambda\mu) \\
Y_cI_c  & yi  & YI \\
\end{array}
\right)
\left(
\begin{array}{cc||c}
(\lambda_c\mu_c) & (10) & (\lambda'\mu') \\
Y_cI_c  & yi  & YI
\end{array}
\right)
\end{eqnarray}
and
\begin{eqnarray}\label{B4}
 \langle \ell S' JJ_3;(\lambda'\mu') Y' I' I'_3|\ell_q^ig^{i8}|\ell S JJ_3(\lambda \mu) Y I I_3\rangle =\delta_{YY'}\delta_{II'}\delta_{I_3I_3'} (-1)^{J+\ell_q+\ell_c-1/2} \nonumber \\
 \times  (2\ell+1) \sqrt{\ell_q(\ell_q+1)(2\ell_q+1)} \sqrt{(2S+1)(2S'+1)} 
 \left\{
\begin{array}{ccc}
 \ell & 1 & \ell \\
 \ell_q & \ell_c  & \ell_q 
\end{array}\right\} 
\left\{
\begin{array}{ccc}
 J & \ell & S \\
 1 & S'  & \ell 
\end{array}\right\} \nonumber \\
\times \sum_{p,p',q,q'} (-1)^{S_c} c^{[N_c-1,1]}_{pp'}(S)c^{[N_c-1,1]}_{q q'}(S')
\left\{
\begin{array}{ccc}
 S' & 1 & S \\
 1/2 & S_c  & 1/2 
\end{array}\right\} \nonumber \\
\times \sum_{Y_c,I_c,y,i}
\frac{3y}{2\sqrt{2}}
\left(
\begin{array}{cc||c}
(\lambda_c\mu_c) & (10) & (\lambda\mu) \\
Y_cI_c  & yi  & YI \\
\end{array}
\right)
\left(
\begin{array}{cc||c}
(\lambda_c\mu_c) & (10) & (\lambda'\mu') \\
Y_cI_c  & yi  & YI
\end{array}
\right).
\end{eqnarray}

To obtain Table \ref{T8} and \ref{b3} we have used the Eqs. (\ref{SU2}) for the coefficients
$c^{[N_c-1,1]}_{pp'}$ and Table IV of Ref. \cite{VERGADOS} 
for the isoscalar factors of SU(3).
From their expressions one can find that all these coefficients and
isoscalar factors 
are of order $N^0_c$.  Then it follows that for states with 
spin and strangeness of order $N^0_c$, the matrix elements of $T^c_8$ are 
of order $N_c$ because $Y_c = Y - y$, $Y= N_c/3 + \mathcal{S}$ so
that $Y_c \sim N_c$. 
 
%%%%%%%%%%%%%%%%%%%%%%%%%%%%%%%%%%%%%%%%%%%%%%%%%%%%%%%%%%%%

\section{}
For completeness here we give the matrix elements of the spin-orbit operator
$O_2$ . They are a generalization from SU(4)
\cite{CCGL} to SU(6) and refer to an excited core with $\ell_c \neq 0$.

\begin{eqnarray}
%\langle \, \1 \negthinspace \: \rangle & =  & 
%\delta_{J'J}\delta_{J'_3J_3}\delta_{L'L}\delta_{I'I}\delta_{I'_3I_3}
%\delta_{S'S}\\
%%%%%%%%%%%%%%%%%%%%%%%%%%%%%%%%%%%%%%%%%%%%%
\langle \ell_q  s \rangle   =   \delta_{J'J}\delta_{J'_3J_3}
\delta_{\lambda' \lambda} \delta_{\mu' \mu} \delta_{Y' Y}
\delta_{I'I}\delta_{I'_3I_3}
(-1)^{J-1/2+\ell_q+\ell_c}
\sqrt{\frac{3}{2} (2S+1)(2S'+1)} \nonumber \\ 
\times 
\sqrt{(2\ell+1)(2\ell'+1)
\ell_q(\ell_q+1)(2\ell_q+1)} %\nonumber \\ 
%& \times & 
 \left\{\begin{array}{ccc}
		\ell & 1 & \ell' \\
		\ell_q & \ell_c & \ell_q
	\end{array}\right\}
\left\{\begin{array}{ccc}
		1 & \ell & \ell' \\
		J & S'& S 
	\end{array}\right\} \nonumber \\
 \times
 \sum_{p p' p''} (-1)^{-S_c} c_{p' p}(S) c_{p'' p}(S')
\left\{\begin{array}{ccc}
		S & 1 & S' \\
		\frac{1}{2} & S_c & \frac{1}{2}
	\end{array}\right\} 
\end{eqnarray}
where $S_c = S - 1/2$ for $p' = 1$ and $S_c = S + 1/2$ for  $p' = 2$
and similarly $S_c = S' - 1/2$ for $p'' = 1$ and $S_c = S' + 1/2$ for $p'' = 2$.
%%%%%%%%%%%%%%%%%%%%%%%%%%%%%%%%%%%%%%%%%%%%%

%%%%%%%%%%%%%%%%%%%%%%%%%%%%%%%%%%%%%%%%%%%%%%%%%%%%%%%%%%%%%%%%%%%%%%
\newpage
{\squeezetable
\begin{table}[htb]
\caption{List of operators and the coefficients resulting from the fit with 
$\chi^2_{\rm dof}  \simeq 1.0$.}
\label{operators}
\renewcommand{\arraystretch}{1.5} % enlarge line spacing
\begin{tabular}{llrrl}
\hline
\hline
Operator & \multicolumn{4}{c}{Fitted coef. (MeV)}\\
\hline
\hline
$O_1 = N_c \ \1 $                                   & \ \ \ $c_1 =  $  & 556 & $\pm$ & 11       \\
$O_2 = \ell_q^i s^i$                                & \ \ \ $c_2 =  $  & -43 & $\pm$ & 47    \\
$O_3 = \frac{3}{N_c}\ell^{(2)ij}_{q}g^{ia}G_c^{ja}$ & \ \ \ $c_3 =  $  & -85 & $\pm$ & 72  \\
$O_4 = \frac{4}{N_c+1} \ell^i t^a G_c^{ia}$         & \ \ \            &     &       &     \\
$O_5 = \frac{1}{N_c}(S_c^iS_c^i+s^iS_c^i)$          & \ \ \ $c_5 =  $  & 253 & $\pm$ & 57  \\
$O_6 = \frac{1}{N_c}t^aT_c^a$                       & \ \ \ $c_6 =  $  & -25 & $\pm$ & 86    \\ 
\hline
$B_1 = t^8-\frac{1}{2\sqrt{3}N_c}O_1$               & \ \ \ $d_1 =  $  & 365 & $\pm$ & 169 \\
$B_2 = T_c^8-\frac{N_c-1}{2\sqrt{3}N_c}O_1$         & \ \ \ $d_2 =  $  &-293 & $\pm$ & 54 \\
$B_4 = 3 \ell^i_q g^{i8}- \frac{\sqrt{3}}{2}O_2$    & \ \ \            &     &       & \vspace{0.2cm}\\
\hline \hline
\end{tabular}
\end{table}}
%%%%%%%%%%%%%%%%%%%%%%%%%%%%%%%%%%%%%%%%%%%%%%%%%%%%%%%%%%%%%%%%%%%%%
{\squeezetable
\begin{table}[htb]
\caption{Matrix elements for octet resonances.}
\label{NUCLEON}
\renewcommand{\arraystretch}{1.5}
\begin{tabular}{lcccccc}
\hline
\hline
   &  \hspace{ .3 cm} $O_1$ \hspace{ .3 cm}  & \hspace{ .3 cm} $O_2$  \hspace{ .3 cm} & \hspace{ .3 cm} $O_3$  \hspace{ .3 cm}&   \hspace{ .3 cm} $O_4$  \hspace{ .3 cm} & \hspace{ .3 cm} $O_5$  \hspace{ .3 cm} & \hspace{ .3 cm} $O_6$  \hspace{ .3 cm} \\
\hline
$^48[{\bf 70},2^+]\frac{7}{2}$  &  $N_c$   &  $\frac{2}{3}$    & $-\frac{3N_c+1}{18N_c}$ & $-\frac{2(3N_c+1)}{9(N_c+1)}$ &  $\frac{5}{2N_c}$ & $\frac{N_c-13}{12N_c}$ \\
$^28[{\bf 70},2^+]\frac{5}{2}$  &  $N_c$   &  $\frac{2(2N_c-3)}{9N_c}$ & 0  & $-\frac{4(N_c+3)(3N_c-2)}{27N_c(N_c+1)}$ &  $\frac{N_c+3}{4N_c^2}$ & $\frac{N_c^2-4N_c-9}{12N_c^2}$ \\
$^48[{\bf 70},2^+]\frac{5}{2}$  &  $N_c$   &  $-\frac{1}{9}$  & $\frac{5(3N_c+1)}{36N_c}$ & $\frac{3N_c+1}{27(N_c+1)}$ &   $\frac{5}{2N_c}$ & $\frac{N_c-13}{12N_c}$\\
$^48[{\bf 70},0^+]\frac{3}{2}$  &  $N_c$   &  0    & 0 &  0 & $\frac{5}{2N_c}$ & $\frac{N_c-13}{12N_c}$\\
$^28[{\bf 70},2^+]\frac{3}{2}$  &  $N_c$   &   $-\frac{2N_c-3}{3N_c}$   & 0 & $\frac{2(N_c+3)(3N_c-2)}{9N_c(N_c+1)}$ & $\frac{N_c+3}{4N_c^2}$ & $\frac{N_c^2-4N_c-9}{12N_c^2}$\\
$^48[{\bf 70},2^+]\frac{3}{2}$  &  $N_c$   &   $-\frac{2}{3}$   & 0 &  $\frac{2(3N_c+1)}{9(N_c+1)}$ & $\frac{5}{2N_c}$ & $\frac{N_c-13}{12N_c}$\\
$^28[{\bf 70},0^+]\frac{1}{2}$  &  $N_c$   &   0   & 0 &  0 & $\frac{N_c+3}{4N_c^2}$ & $\frac{N_c^2-4N_c-9}{12N_c^2}$\\
$^48[{\bf 70},2^+]\frac{1}{2}$  &  $N_c$   &    $-1$   &  $-\frac{7(3N_c+1)}{36N_c}$ & $\frac{3N_c+1}{3(N_c+1)}$&  $\frac{5}{2N_c}$ & $\frac{N_c-13}{12N_c}$\vspace{0.2cm} \\
\hline
\hline
\end{tabular}
\end{table}}
%%%%%%%%%%%%%%%%%%%%%%%%%%%%%%%%%%%%%%%%%%%%%%%%%%%%%%%%%%%%%%%%%%%%

{\squeezetable
\begin{table}[htb]
\caption{Matrix elements for decuplet resonances.}
\label{DELTA}
\renewcommand{\arraystretch}{1.5}
\begin{tabular}{lcccccc}
\hline
\hline
   &  \hspace{ .3 cm} $O_1$ \hspace{ .3 cm}  & \hspace{ .3 cm} $O_2$  \hspace{ .3 cm} &  \hspace{ .3 cm} $O_3$  \hspace{ .3 cm} & \hspace{ .3 cm} $O_4$  \hspace{ .3 cm} & \hspace{ .3 cm} $O_5$  \hspace{ .3 cm} & \hspace{ .3 cm} $O_6$  \hspace{ .3 cm} \\
\hline

$^210[{\bf 70},2^+]\frac{5}{2}$  &   $N_c$   &  $-\frac{2}{9}$ & 0  & $\frac{2(3N_c+7)}{27(N_c+1)}$ & $\frac{1}{N_c}$ & $\frac{N_c+5}{12N_c}$\\
$^210[{\bf 70},2^+]\frac{3}{2}$  &   $N_c$   &  $\frac{1}{3}$  & 0  & $-\frac{3N_c+7}{9(N_c+1)}$ & $\frac{1}{N_c}$ & $\frac{N_c+5}{12N_c}$\\
$^210[{\bf 70},0^+]\frac{1}{2}$  &   $N_c$   &  0 & 0 & 0 & $\frac{1}{N_c}$ & $\frac{N_c+5}{12N_c}$\vspace{0.2cm}\\
\hline
\hline
\end{tabular}
\end{table}}

{\squeezetable
\begin{table}
\caption{Matrix elements for singlet resonances.}
\renewcommand{\arraystretch}{1.5}
\label{SINGLET}
\begin{tabular}{lcccccc}
\hline
\hline
   &  \hspace{ .3 cm} $O_1$ \hspace{ .3 cm}  & \hspace{ .3 cm} $O_2$  \hspace{ .3 cm} &  \hspace{ .3 cm} $O_3$  \hspace{ .3 cm} & \hspace{ .3 cm} $O_4$  \hspace{ .3 cm} & \hspace{ .3 cm} $O_5$  \hspace{ .3 cm} & \hspace{ .3 cm} $O_6$  \hspace{ .3 cm} \\
\hline
$^21[{\bf 70},2^+]\frac{5}{2}$  &  $N_c$   &   $\frac{2}{3}$   &   0 & 0  & 0 & $-\frac{N_c+5}{6N_c}$\\
$^21[{\bf 70},2^+]\frac{3}{2}$  &  $N_c$   &   $-1$   &   0  & 0 & 0 & $-\frac{N_c+5}{6N_c}$\\
$^21[{\bf 70},0^+]\frac{1}{2}$  &  $N_c$   &   0  &   0   & 0 & 0 & $-\frac{N_c+5}{6N_c}$\vspace{0.2cm}\\
\hline
\hline
\end{tabular}
\end{table}}

%%%%%%%%%%%%%%%%%%%%%%%%%%%%%%%%%%%%%%%%%%%%%%%%%%%%%%%%%%%%%%%%%%%%%%%%%%%

\begin{turnpage}
\renewcommand{\arraystretch}{1.5}
{\squeezetable
\begin{table}
\caption{Matrix elements of $t_8$ and $T^c_8$ as a function of $N_c$, the
isospin $I$ and the strangeness $\mathcal{S}$. The off-diagonal matrix elements have $(\mathcal{S}=-1,I=1)$ or $(\mathcal{S}=-2,I=1/2)$ for $^28_J- {}^210_J$ and $(\mathcal{S}=0,I=0)$ for $^28_J- {}^21_J$.}
\label{T8}
\begin{tabular}{ccc}
\hline \hline
             &     $t_8$      &        $T^c_8$ \\
\hline
$^28_J$ & $\frac{N_c^3+[\mathcal{S}(5-\mathcal{S})+4I(I+1)-1]N_c^2-3[\mathcal{S}(2-\mathcal{S})+4I(I+1)-2]N_c+9\mathcal{S}}{2\sqrt{3}N_c(N_c-1)(N_c+3)}$ & $\frac{N_c^4+(3\mathcal{S}+1)N_c^3+[(\mathcal{S}(\mathcal{S}+1)-4I(I+1)-2]N_c^2-3[\mathcal{S}(\mathcal{S}+1)-4I(I+1)+2]N_c-9\mathcal{S}}{2\sqrt{3}N_c(N_c-1)(N_c+3)}$ \\
$^48_J$ & $\frac{2N_c-4I(I+1)+\mathcal{S}(\mathcal{S}+4)+1}{4\sqrt{3}(N_c-1)}$ & $\frac{2N_c^2+2(3\mathcal{S}-2)N_c+4I(I+1)-\mathcal{S}(\mathcal{S}+10)-1}{4\sqrt{3}(N_c-1)}$ \\
$^210_J$ & $\frac{2N_c+4I(I+1)-\mathcal{S}(\mathcal{S}-8)-5}{4\sqrt{3}(N_c+5)}$ & $\frac{2N_c^2+2(3\mathcal{S}+4)N_c-4I(I+1)+\mathcal{S}(\mathcal{S}+22)+5}{4\sqrt{3}(N_c+5)}$ \\
$^21_J$ & $\frac{-2N_c^2-2(3\mathcal{S}+1)N_c+12I(I+1)-3\mathcal{S}(\mathcal{S}+2)+3}{2\sqrt{3}(N_c+1)(N_c+3)}$ & $\frac{N_c^3+3(\mathcal{S}+2)N_c^2+(18\mathcal{S}+5)N_c-12I(I+1)+3\mathcal{S}(\mathcal{S}+5)-3}{2\sqrt{3}(N_c+1)(N_c+3)}$\\
$^28_J- {}^210_J$  & $\sqrt{\frac{2}{3}}\sqrt{\frac{N_c+3}{N_c(N_c-1)(N_c+5)}}$ &  $-\sqrt{\frac{2}{3}}\sqrt{\frac{N_c+3}{N_c(N_c-1)(N_c+5)}}$  \\
$^28_J- {}^21_J$ & $\frac{3(N_c-1)}{2\sqrt{N_c}(N_c+3)}$ &$-\frac{3(N_c-1)}{2\sqrt{N_c}(N_c+3)}$
\vspace{0.2cm} \\
\hline
\hline
\end{tabular}
\end{table}}
\end{turnpage}

{\squeezetable
\begin{table}
\caption{Matrix elements of the term $3\ell^i g^{i8}$ of $B_4$.}\label{b3}
\renewcommand{\arraystretch}{1.5}
\begin{tabular}{cc}
 \hline
 \hline
  & $3\ell^ig^{i8}$ \\ \hline
 $^48[{\bf 70},2^+]\frac{7}{2}$ & $\frac{2N_c-4I(I+1)+\mathcal{S}(\mathcal{S}+4)+1}{2\sqrt{3}(N_c-1)}$ \\
 $^28[{\bf 70},2^+]\frac{5}{2}$  & $\frac{4N_c^3+4I(I+1)(9+N_c(7N_c-12))-9(\mathcal{S}-1)^2-N_c^2(\mathcal{S}-1)(7\mathcal{S}-19)+12N_c(\mathcal{S}(\mathcal{S}-5)+1)}{6\sqrt{3}N_c(N_c-1)(N_c+3)}$ \\
 $^48[{\bf 70},2^+]\frac{5}{2}$ & $-\frac{2N_c-4I(I+1)+\mathcal{S}(\mathcal{S}+4)+1}{12\sqrt{3}(N_c-1)}$ \\
 $^48[{\bf 70},0^+]\frac{3}{2}$ & 0 \\
 $^28[{\bf 70},2^+]\frac{3}{2}$ & $-\frac{4N_c^3+4I(I+1)(9+N_c(7N_c-12))-9(\mathcal{S}-1)^2-N_c^2(\mathcal{S}-1)(7\mathcal{S}-19)+12N_c(\mathcal{S}(\mathcal{S}-5)+1)}{4\sqrt{3}N_c(N_c-1)(N_c+3)}$ \\
 $^48[{\bf 70},2^+]\frac{3}{2}$ & $-\frac{2N_c-4I(I+1)+\mathcal{S}(\mathcal{S}+4)+1}{2\sqrt{3}(N_c-1)}$ \\
 $^28[{\bf 70},0^+]\frac{1}{2}$ & 0 \\ 
 $^48[{\bf 70},2^+]\frac{1}{2}$ &  $-\frac{\sqrt{3}(2N_c-4I(I+1)+\mathcal{S}(\mathcal{S}+4)+1)}{4(N_c-1)}$ \\
 \hline
 $^210[{\bf 70},2^+]\frac{5}{2}$  & $-\frac{2N_c+4I(I+1)-\mathcal{S}(\mathcal{S}-8)-5}{6\sqrt{3}(N_c+5)}$ \\
 $^210[{\bf 70},2^+]\frac{3}{2}$  & $\frac{2N_c+4I(I+1)-\mathcal{S}(\mathcal{S}-8)-5}{6\sqrt{3}(N_c+5)}$ \\
 $^210[{\bf 70},0^+]\frac{1}{2}$  & 0 \\
 \hline
 $^21[{\bf 70},2^+]\frac{5}{2}$  & $\frac{3+12I(I+1)-2N_c(N_c+1)-3\mathcal{S}(\mathcal{S}+2N_c+2)}{\sqrt{3}(N_c+1)(N_c+3)}$ \\
 $^21[{\bf 70},2^+]\frac{3}{2}$  & $-\frac{\sqrt{3}(3+12I(I+1)-2N_c(N_c+1)-3\mathcal{S}(\mathcal{S}+2N_c+2))}{2(N_c+1)(N_c+3)}$ \\ 
 $^21[{\bf 70},0^+]\frac{1}{2}$  & 0 \\
 \hline \hline
\end{tabular}
\end{table}

{\squeezetable
\begin{table}
%\begin{sidewaystable}
\caption{The partial contribution and the total mass (MeV) predicted by the $1/N_c$ expansion. The last two columns give  the empirically known masses.}\label{MASSES}
\renewcommand{\arraystretch}{1.5}
\begin{tabular}{crrrrrrrcccl}\hline \hline
                    &      \multicolumn{7}{c}{Part. contrib. (MeV)}  & \hspace{.0cm} Total (MeV)   & \hspace{.0cm}  Exp. (MeV)\hspace{0.0cm}& &\hspace{0.cm}  Name, status \hspace{.0cm} \\

\cline{2-8}
                    &   \hspace{.0cm}   $c_1O_1$  & \hspace{.0cm}  $c_2O_2$ & \hspace{.0cm}$c_3O_3$ &\hspace{.0cm}  $c_5O_5$ &\hspace{.0cm}  $c_6O_6$ & $d_1B_1$& $d_2B_2$&    &        \\
\hline
$^4N[{\bf 70},2^+]\frac{7}{2}$        & 1667 & -29 & 16 & 211 & 7 & 0   & 0     &   $1872\pm 46$  & $2016\pm104$ & & $F_{17}(1990)$**  \\
$^4\Lambda[{\bf 70},2^+]\frac{7}{2}$  &      &     &    &     &   &  0   & 254  &   $2125\pm 72$  & $2094\pm78$  & & $F_{07}(2020)$* \\
$^4\Sigma[{\bf 70},2^+]\frac{7}{2}$   &      &     &    &     &   & -211 & 85  &   $1745\pm95$  &              & & \\
$^4\Xi[{\bf 70},2^+]\frac{7}{2}$      &      &     &    &     &   &  -105 & 423 &   $2189\pm81$   &              & & \vspace{0.2cm}\\
\hline
$^2N[{\bf 70},2^+]\frac{5}{2}$        & 1667 & -10 & 0  & 42  & 3 &  0  &  0    &   $1703\pm29$   &              & & \\
$^2\Lambda[{\bf 70},2^+]\frac{5}{2}$  &      &     &    &     &   & -105   &  169 & $1766\pm26$   &              & & \\
$^2\Sigma[{\bf 70},2^+]\frac{5}{2}$   &      &     &    &     &   & -105   &  169 & $1766\pm26$   &              & & \\
$^2\Xi[{\bf 70},2^+]\frac{5}{2}$      &      &     &    &     &   & -211   &  338 & $1830\pm58$  &              & & \vspace{0.2cm}\\
\hline
$^4N[{\bf 70},2^+]\frac{5}{2}$        & 1667 &  5  &-39 & 211 & 7 &  0  &  0    &   $1850\pm44$   & $1981\pm200$ & & $F_{15}(2000)$** \\
$^4\Lambda[{\bf 70},2^+]\frac{5}{2}$  &      &     &    &     &   &  0    & 254 &  $2104\pm39$   & $2112\pm40$  & & $F_{05}(2110)$*** \\
$^4\Sigma[{\bf 70},2^+]\frac{5}{2}$   &      &     &    &     &   &  -211   & 85 &  $1724\pm111$  &              & & \\
$^4\Xi[{\bf 70},2^+]\frac{5}{2}$      &      &     &    &     &   &  -105   &  423 & $2167\pm54$   &              & & \vspace{0.2cm}\\
\hline
$^4N[{\bf 70},0^+]\frac{3}{2}$        & 1667 &  0  &  0 & 211 & 7 &  0  &  0    &   $1885\pm17$   & $1879\pm17$  & & $P_{13}(1900)$** \\
$^4\Lambda[{\bf 70},0^+]\frac{3}{2}$  &      &     &    &     &   &  0     &  254 &  $2138\pm42$   &              & & \\
$^4\Sigma[{\bf 70},0^+]\frac{3}{2}$   &      &     &    &     &   & -211   &  85  & $1758\pm100$  &              & & \\
$^4\Xi[{\bf 70},0^+]\frac{3}{2}$      &      &     &    &     &   & -105   &  423 &
 $2202\pm56$   &              & & \vspace{0.2cm}\\
\hline
$^2N[{\bf 70},2^+]\frac{3}{2}$        & 1667 & 14  &  0 & 42  & 3 &  0  &  0    &   $1727\pm31$   &              & & \\
$^2\Lambda[{\bf 70},2^+]\frac{3}{2}$  &      &     &    &     &   &  -105   & 169 &   $1790\pm29$   &              & & \\
$^2\Sigma[{\bf 70},2^+]\frac{3}{2}$   &      &     &    &     &   & -105   &  169 & $1790\pm29$   &              & & \\
$^2\Xi[{\bf 70},2^+]\frac{3}{2}$      &      &     &    &     &   & -211   &  338 & $1854\pm59$  &              & & \vspace{0.2cm}\\
\hline
$^4N[{\bf 70},2^+]\frac{3}{2}$        & 1667 & 29  &  0 & 211 & 7 &  0  &  0    &   $1914\pm33$   &              & & \\
$^4\Lambda[{\bf 70},2^+]\frac{3}{2}$  &      &     &    &     &   &   0    &  254 & $2167\pm41$   &              & & \\
$^4\Sigma[{\bf 70},2^+]\frac{3}{2}$   &      &     &    &     &   & -211  &   85 & $1787\pm103$  &              & & \\
$^4\Xi[{\bf 70},2^+]\frac{3}{2}$      &      &     &    &     &   & -105   &  423 & $2231\pm56$   &              & &
\vspace{0.2cm} \\
\hline
\hline
\end{tabular}
%\end{sidewaystable}
\end{table}}
%%%%%%%%%%%%%%%%%%%%%%%%%%%%%%%%%%%%%%%%%%%%%%%%%%%%%%%%%%%%%%%%%%%%%%%%%%%

{\squeezetable
\begin{table}
\label{multiplet}
\renewcommand{\arraystretch}{1.5}
\begin{tabular}{crrrrrrrcccl}\hline \hline
                    &      \multicolumn{7}{c}{Part. contrib. (MeV)}  & \hspace{.0cm} Total (MeV)   & \hspace{.0cm}  Exp. (MeV)\hspace{0.0cm}& &\hspace{-0.2cm}  Name, status \hspace{.0cm} \\

\cline{2-8}
                    &   \hspace{.0cm}   $c_1O_1$  & \hspace{.0cm}  $c_2O_2$ &
		    \hspace{.0cm}$c_3O_3$ &\hspace{.0cm}  $c_5O_5$  &\hspace{.0cm}  $c_6O_6$& $d_1B_1$& $d_2B_2$&    &        \\
\hline
$^2N[{\bf 70},0^+]\frac{1}{2}$        &  1667&  0  &  0 & 42  & 3  &  0  &  0    &   $1712\pm27$   &  $1710\pm30$ & & $P_{11}(1710)$*** \\
$^2\Lambda[{\bf 70},0^+]\frac{1}{2}$  &      &     &    &     &    & -105   & 169 &  $1776\pm24$   &            & & \\
$^2\Sigma[{\bf 70},0^+]\frac{1}{2}$   &      &     &    &     &    &  -105   & 169 &  $1776\pm24$   &  $1760\pm27$            & & $P_{11}(1770)$* \\ 
$^2\Xi[{\bf 70},0^+]\frac{1}{2}$      &      &     &    &     &    & -211   &  338 & $1839\pm57$  &              & & \vspace{0.2cm}\\
\hline
$^4N[{\bf 70},2^+]\frac{1}{2}$        & 1667 & 43  & 55 & 211 &  7 &  0  &  0    &   $1983\pm26$   &  $1986\pm26$ & & $P_{11}(2100)$* \\
$^4\Lambda[{\bf 70},2^+]\frac{1}{2}$  &      &     &    &     &    &  0  & 254   &   $2237\pm57$   &              & & \\
$^4\Sigma[{\bf 70},2^+]\frac{1}{2}$   &      &     &    &     &    & -211 & 85  &   $1857\pm90$  &              & & \\
$^4\Xi[{\bf 70},2^+]\frac{1}{2}$      &      &     &    &     &    & -105  & 423 &   $2301\pm68$   &              & & \vspace{0.2cm}\\
\hline
$^2\Delta[{\bf 70},2^+]\frac{5}{2}$   & 1667 & 10  & 0  & 84  & -6 &  0  &  0    &
$1756\pm32$   & $1976\pm237$ & & $F_{35}(2000)$**\\
$^2\Sigma'[{\bf 70},2^+]\frac{5}{2}$  &      &     &    &     &    & -105 & 169   &   $1819\pm46$   &              & & \\
$^2\Xi'[{\bf 70},2^+]\frac{5}{2}$     &      &     &    &     &    & -211 & 338  &   $1883\pm77$  &              & & \\
$^2\Omega[{\bf 70},2^+]\frac{5}{2}$   &      &     &    &     &    & -316 & 507 &   $1946\pm113$  &              & &\vspace{0.2cm} \\
\hline
$^2\Delta[{\bf 70},2^+]\frac{3}{2}$   & 1667 & -14 & 0  & 84  & -6 &  0  & 0   & $1731\pm35$ &                &  & \\
$^2\Sigma'[{\bf 70},2^+]\frac{3}{2}$  &      &     &    &     &    & -105 & 169  & $1795\pm48$ &                &  & \\
$^2\Xi'[{\bf 70},2^+]\frac{3}{2}$     &      &     &    &     &    & -211 & 338 & $1859\pm78$&                &  & \\
$^2\Omega[{\bf 70},2^+]\frac{3}{2}$   &      &     &    &     &    & -316  & 507 & $1922\pm113$&                &  & \vspace{0.2cm}\\
\hline
$^2\Delta[{\bf 70},0^+]\frac{1}{2}$   & 1667 &  0  &  0 & 84  & -6 & 0  &  0  & $1746\pm31$ &  $1744\pm36$   &  & $P_{31}(1750)$* \\
$^2\Sigma'[{\bf 70},0^+]\frac{1}{2}$  &      &     &    &     &    & -105  & 169 & $1810\pm45$ &  $1896\pm95$   &  & $P_{11}(1880)$** \\
$^2\Xi'[{\bf 70},0^+]\frac{1}{2}$     &      &     &    &     &    & 211 & 338 & $1873\pm77$                 &  & \\
$^2\Omega[{\bf 70},0^+]\frac{1}{2}$   &      &     &    &     &    & 316& 507 & $1937\pm112$&                &  & \vspace{0.2cm}\\
\hline
$^2\Lambda'[{\bf 70},2^+]\frac{5}{2}$ & 1667 & -29 &  0 &  0  & 11 & -105 & 169 & $1713\pm51$ &                &  &\vspace{0.2cm} \\
\hline
$^2\Lambda'[{\bf 70},2^+]\frac{3}{2}$ & 1667 &  43 &  0 &  0  & 11 & -105 & 169 & $1785\pm62$ &                &  & \vspace{0.2cm}\\
\hline
$^2\Lambda'[{\bf 70},0^+]\frac{1}{2}$ & 1667 &  0  &  0 &  0  & 11 & -105& 169 & $1742\pm40$ &  $1791\pm64$   &  & $P_{01}(1810)$*** \vspace{0.2cm}\\
\hline
\hline
\end{tabular}
%\end{sidewaystable}
\end{table}}

%%%%%%%%%%%%%%%%%%%%%%%%%%%%%%%%%%%%%%%%%%%%%%%%%%%%%%%%%%%%%%%%%%%%%%
\end{document}